\documentclass[twocolumn,floatfix,aps,prc,showpacs]{revtex4}
 
\usepackage[final]{epsfig}
\usepackage{amsmath}
\usepackage{bm}
\usepackage{graphicx}

\usepackage{color}

\begin{document}
 
\title{Pathlength dependence of energy loss within in-medium showers}
 
\author{Thorsten Renk}
\email{thorsten.i.renk@jyu.fi}

\affiliation{Department of Physics, P.O. Box 35, FI-40014 University of 
             Jyv\"askyl\"a, Finland}
\affiliation{Helsinki Institute of Physics, P.O. Box 64, FI-00014 University 
             of Helsinki, Finland}


\begin{abstract}
Studying the pathlength dependence of high $P_T$ hadron suppression in heavy-ion collisions by measuring the dependence of hard hadron production on the angle $\phi$ with the reaction plane in non-central collisions has so far been one of the most successful tools in constraining the microscopical picture of leading parton energy loss. With the imminent start of the LHC heavy-ion program and the possibility of full jet reconstruction, the focus has shifted to models which are capable of simulating full in-medium parton showers rather than tracing the leading parton only. Yet, on the level of single inclusive hadron observables, such shower models need to reproduce the findings of leading parton energy loss models, in particular they need to reproduce the correct reaction plane angle dependence of hadron suppression. The aim of this work is to illustrate at the example of the in-medium shower code YaJEM (Yet another Jet Energy-loss Model) how pathlength dependence arises in a MC shower formulation, how it relates to pathlength dependence of leading parton energy loss and how model results compare with data on the reaction plane dependence of the nuclear suppression factor $R_{AA}(\phi)$.

\end{abstract}
 
\date{\today}

\maketitle

\section{Introduction}

The idea to use the modification of high $P_T$ hadron production due to the bulk medium produced in ultrarelativistic heavy-ion collisions to study properties of the medium is known as 'jet tomography' \cite{Tomo1,Tomo2,Tomo3,Tomo4,Tomo5}. However, in order to gain any tomographical information about the macroscopical medium density distribution and its evolution from measured data in practice, it is necessary to know first the microscopical mechanism by which the medium modifies high $P_T$ hadron spectra. A less ambitious approach is to determine under what assumptions a combination of macroscopical evolution in terms of a bulk evolution model like fluid dynamics and microscopical parton-medium interaction model can describe a large body of data in a systematic way \cite{SysJet1,SysJet2,SysHyd1,SysHyd2}. So far, most of these studies have been carried out for models which describe the medium modification as energy loss from the leading shower parton prior to hadronization.

An emerging result is that the pathlength dependence of energy loss, accessible experimentally e.g. via the dependence of the nuclear suppression factor $R_{AA}(\phi)$ on angle of outgoing hadrons with the reaction plane $\phi$ provides a crucial constraint for the microscopical physics. In particular, the pathlength dependence cannot be linear in a constant medium (as characteristic for elastic or generally incoherent processes) \cite{Elastic1,Elastic2} but must be at least quadratic \cite{SysHyd3} characteristic of coherence, possibly even cubic \cite{StrongCoupling,SysHyd3} as found in some versions of strongly coupled QCD-like models. 

On the other hand, going beyond the capability to describe single inclusive hadron observables to full jets, Monte-Carlo (MC) models capable of describing the development of a parton shower originating from a highly virtual initial parton inside a medium, followed by hadronization into jets are being developed. Both elastic collisions \cite{JEWEL} as well as medium-induced radiation \cite{YaJEM1,YaJEM2,QPYTHIA,MARTINI} have been considered as mechanisms for the modification. 

Monte-Carlo (MC) models are not natural frameworks to incorporate the physics of quantum coherence (which appears necessary to reproduce the pathlength dependence of energy loss seen in the data), but it has been demonstrated in \cite{JEWEL-LPM} that an algorithm can be devised which consistently reproduces known eikonal infinite energy analytical limits. Yet in the same work, it has also been shown that including realistic effects such as finite energy corrections, the pathlength dependence is effectively weaker than quadratic (see also \cite{YaJEM-proc}). This is in manifest contradiction to systematic investigations which show that the data for $R_{AA}(\phi)$ are marginally in agreement with a quadratic dependence and would be easier reproduced with a cubic dependence \cite{SysHyd3}.

The aim of this paper is to illustrate the connection between the pathlength dependence of energy loss in leading parton models and the effect of pathlength on an in-medium shower more clearly and to suggest a possible resolution for the apparent puzzle outlined above by suggesting that the pathlength dependence observed in the data might be due to a yet different effect. For all computations in the following, the radiative energy loss scenario (RAD) of the MC in-medium shower code YaJEM is used.

\section{The model}

\subsection{General description}

YaJEM is described in detail in \cite{YaJEM1,YaJEM2}, here we just reproduce the essential details necessary to understand how  pathlength dependence arises within the model.

In the absence of a medium, YaJEM utilizes the PHYSHOW algorithm \cite{PYSHOW} in momentum space to model the evolution from a highly virtual initial parton to a shower of low virtuality partons and the Lund model \cite{Lund} to hadronize this shower. In this framework, the QCD evolution from a highly virtual state to a parton shower is modelled as a series of branchings $a \rightarrow bc$ where the virtuality scale decreases at each step.

In the presence of a medium, it is no longer sufficient to model in momentum space only since the medium evolution is caused by pressure gradients in spacetime and the density of matter felt by hard partons is a function of position space rather than momentum space. In order to make a link between the shower evolution in momentum space and the medium density evolution in position space, YaJEM utilizes the uncertainty relation to compute the average formation time of a virtual daughter parton $b$ generated in the branching of a parent $a$ as 

\begin{equation}
\label{E-Lifetime}
\langle \tau_b \rangle = \frac{E_b}{Q_b^2} - \frac{E_b}{Q_a^2}.
\end{equation}  

where $E_i$ denote parton energies and $Q_i$ parton virtualities. YaJEM provides the additional option not to use the average formation time but rather for each branching an actual formation time given by an exponential distribution 

\begin{equation}
\label{E-RLifetime}
P(\tau_b) = \exp\left[- \frac{\tau_b}{\langle \tau_b \rangle}  \right]
\end{equation}

which averages to $\langle \tau_b \rangle$. This allows to specify the temporal structure of the shower, and combined with the assumption that all shower partons sample the medium along the eikonal trajectory of the shower-initiating parton, the complete space-time picture of the shower is fixed.

In the radiative energy loss scenario (RAD) of YaJEM, the parton-medium interaction is modelled by a medium-induced virtuality gain during the parton formation time

\begin{equation}
\label{E-Qgain}
\Delta Q_a^2 = \int_{\tau_a^0}^{\tau_a^0 + \tau_a} d\zeta \hat{q}(\zeta)
\end{equation}

where the time $\tau_a$ is given by Eq.~(\ref{E-RLifetime}), the time $\tau_a^0$ is known in the simulation as the endpoint of the previous branching process and the integration $d\zeta$ is along the eikonal trajectory of the shower-initiating parton. If the parton is a gluon, the virtuality transfer from the medium is increased by the ratio of their Casimir color factors, $3/\frac{4}{3} =  2.25$. The medium is characterized via the parameter $\hat{q}(\zeta)$ where a relation like

\begin{equation}
\label{E-qhat}
\hat{q}(\zeta) = K \cdot 2 \cdot [\epsilon(\zeta)]^{3/4} (\cosh \rho(\zeta) - \sinh \rho(\zeta) \cos\psi)
\end{equation}

is assumed as the link to fluid dynamical parameters such as the energy density $\epsilon$ or the local transverse flow rapidity $\rho$ (at angle $\psi$ with the parton trajectory).

One will observe that Eq.~(\ref{E-Lifetime}),(\ref{E-Qgain}) are not independent --- parton virtuality determines the parton formation time, but the formation time in turn determines the virtuality gain from the medium. If $\Delta Q_a^2 \ll Q_a^2$, holds, i.e. the virtuality picked up from the medium is a correction to the initial parton virtuality, we may add $\Delta Q_a^2$ to the virtuality of parton $a$ before determining the kinematics of the next branching. If the condition is not fulfilled, the lifetime is determined by $Q_a^2 + \Delta Q_a^2$ and may be significantly shortened by virtuality picked up from the medium. In this case we iterate Eqs.~(\ref{E-Lifetime}),(\ref{E-Qgain}) to determine a self-consistent pair of $(\langle \tau_a\rangle, \Delta Q^2_a)$. This ensures that on the level of averages, the lifetime is treated consistently with the virtuality picked up from the medium. The actual lifetime is still determined by Eq.~(\ref{E-RLifetime}). This consistency condition is fundamentally the same as the algorithm presented in \cite{JEWEL-LPM}. It corresponds to the limit of a coherent summation of a large number of medium interactions during the formation time (in YaJEM, virtuality gain is a smooth function of pathlength, rather than a sum over discrete interactions transferring virtuality) and hence represents LPM interference in this limit.

However, for technical reasons, despite the consistency condition for medium-induced virtuality gain, the in-medium shower simulated by YaJEM remains an expansion for thin media. The reason is that, as in the PYSHOW algorithm, parton  splitting no longer occurs once a parton has reached a lower virtuality scale $Q_0$ (where $Q_0 = O(1)$ GeV --- to be more precise, for massive partons the limit is given by $m^q_{eff} + m^g_{eff}$ with $m^g_{eff} = \frac{1}{2} Q_0$ and $m^q_{eff} = \sqrt{m_q^2 + \frac{1}{4}Q_0^2}$ and $m_q$ the current quark mass). Partons falling at any point of the evolution below this scale are set on-shell and do not longer branch. YaJEM inherits this condition also in a medium. This limits the distance scale $L_{max}$ for which an in-medium shower is evolved reliably parametrically to $L_{max} \sim E/Q_0^2$ where $E$ is the energy of the shower-initiating parton.

\subsection{Pathlength dependence}

It should now be clear from where any dependence of parton distributions created by YaJEM on the in-medium pathlength arise.

\begin{itemize}
\item
The leading effect is given by Eq.~(\ref{E-Qgain}) which states that the amount of medium-induced virtuality is a monotonically increasing function of the pathlength. If $\hat{q}$ is a constant, this states that the total amount of medium-induced virtuality (or the number of scattering centers probed by a parton) is linearly proportional to the pathlength. Since Eq.~(\ref{E-Qgain}) holds for {\em every} parton in the shower, this would lead to a crescendo-like scenario \cite{Crescendo} for the virtuality transfer to the shower as a whole in the absence of finite energy or finite virtuality corrections (in practice however, such corrections turn out to dominate the dynamics \cite{Bryon}).

\item
The consistency condition between formation time and medium-induced virtuality forms a second important ingredient. It states that low $Q^2$, i.e. collinear medium-induced radiation may not occur after arbitrarily short formation times. When applied to a single almost-on-shell parton propagating through a constant medium, it gives rise to a quadratic pathlength dependence of mean radiated energy as in \cite{JEWEL-LPM}. However, note that since YaJEM requires $Q > Q_0$, this situation is strictly never realized. To some degree, its effect is also obscured by the randomization of formation times Eq.~(\ref{E-RLifetime}) as will be demonstrated later.

\item
The minimum virtuality scale $Q_0$ down to which partons are evolved induces a spurious length dependence whenever a pathlength $L > L_{max}$ is probed --- for $L \gg L_{max}$ (i.e. far outside the validity region of the model), all partons in the shower are on-shell and no branching occurs for any density of the medium. Thus, when applied outside the region of validity, YaJEM leads to {\em less} modification of showers than expected in reality.

\item
An interesting proposal, suggested in \cite{Abhijit1,Abhijit2}, is to {\em define} the minimum virtuality scale by

\begin{equation}
\label{E-Q0}
Q_0 = \sqrt{E/L}
\end{equation}

where $L$ is the actual pathlength a shower is evolved in the medium and $E$ the energy of the shower-initiating parton, rather than leave it a constant. If this is done, it replaces the spurious $Q_0$ dependence by a genuine scale dependence of the shower evolution which states that the medium cannot induce radiation below a scale $Q_0$. We will test this prescription in the following.

\end{itemize}

In practical terms, it was found in \cite{YaJEM1,YaJEM2} that many observables (e.g. the fragmentation function) follow to good approximation a scaling law with the integrated total virtuality gain from the medium, $\Delta Q^2 = \int_0^L d\xi \hat{q} (\xi)$ inside the region of validity. This implies that the pathlength dependence of such observables is approximately linear in a constant medium.

\subsection{Implications of a dynamical minimum virtuality}

Let us take some time to consider the implications of Eq.~(\ref{E-Q0}) in some more detail. The obvious problem is that $Q_0$ regulates both the endpoint for the vacuum and for the in-medium parton shower evolution (which can't be disentangled on a branching-by-branching basis). While length dependence in the medium contribution is reasonable, there is no justification for any length effect on the vacuum shower evolution. In contrast, in vacuum, $Q_0$ regulates the transition between the perturbative partonic and non-perturbative hadronic domain of fragmentation. This in turn is an ill-defined condition in a deconfined medium, as hadrons cannot be formed in such a medium.

 Eq.~(\ref{E-Q0}) can therefore not be useful in general but must  be seen as an approximation for a situation in which the shower evolution is dominated by medium effects and in which the kinematics is such that the possibility of hadrons being formed in the medium is not a crucial issue. The latter condition is readily fulfilled when considering single inclusive high $P_T$ pion observables (as done in this work), as the pion formation length (estimated as $\tau \sim E_\pi/m_{\pi}^2$) is far larger than the medium size. Note that this is not the case for the low $z$ part of the fragmentation function or for the production of heavier hadrons.

\begin{figure}[htb]
\begin{center}
\epsfig{file= 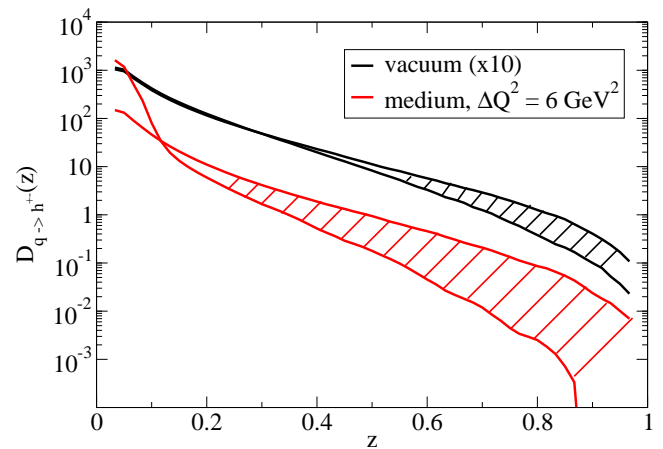, width=8.5cm}
\end{center}
\caption{\label{F-FFscale}(Color online) Response of the computed fragmentation function of a 20 GeV quark into charged hadrons to choices of the minimum virtuality scale $Q_0^2$ from 0.5 GeV$^2$ to 2 GeV$^2$, both for the vacuum case and for a medium-modified shower for total medium induced firtuality $\Delta Q^2 = 6$ GeV$^2$ as typical for the majority of partons in RHIC high $P_T$ observables within YaJEM.}
\end{figure}

It turns out that the first condition is also fulfilled for typical in-medium pathlengths as relevant for computing observables. Fig.\ref{F-FFscale} shows vacuum and medium-modified  fragmentation functions as bands indicating their response to a change in $Q_0^2$ from 0.5 GeV$^2$ to 2 GeV$^2$ for typical parton paths at RHIC kinematics. It is readily apparent from the figure that the majority of the variation with $Q_0$ seen in the in-medium case cannot be accounted for by the comparatively small effect of the scale change in vacuum, indicating that the pathlength condition in this situation indeed dominates over the scale dependence of the transition to the non-perturbative regime. Part of the reason is that $Q_0$ does not only define the endpoint of the perturbative shower description, but also the starting point of the non-perturbative Lund hadronization model. Thus, to some degree, in the vacuum decreased showering is compensated by an increased action of the Lund model, whereas no such compensation occurs in the medium case (since the hadronization is by assumption always treated as in vacuum).

Nevertheless, Eq.~(\ref{E-Q0}) has to be regarded as a first approximation. Ultimately, a criterion which also takes into account the virtuality $\Delta Q^2$ induced by the medium along the parton path would be superior. This can be understood in the limit of small pathlength $L$. For applications to observables, $\Delta Q^2$ and $L$ are not independent and small $L$ also implies small $\Delta Q^2$. In this case, the appropriate limit for $Q_0$ is the vacuum default value $Q_0 = 1$ GeV rather than  the value given by Eq.~(\ref{E-Q0}), as the assumption that the medium effect dominates the dynamics of the shower is not fulfilled when $\Delta Q^2$ is small. However, one could think of an artificial problem in which $\Delta Q^2$ is large while $L$ is small, in which case the limit for small pathlengths is not the vacuum.

For the time being we will nevertheless explore the consequences of the approximation Eq.~(\ref{E-Q0}) as it stands rather than introduce a more general prescription which brings additional model dependence. This means that the results are not reliable for small pathlengths where the shower is not dominated by the medium, but the underlying assumption is justified by the vast majority of partons encountered in the computation of experimentally relevant observables.

\subsection{Medium-modified showers and energy loss}

It is not straightforward for medium-modified shower models to make contact with leading parton energy loss models. Typically in an energy loss model like \cite{Tomo1,Tomo2,Tomo3} there is a clear distinction between the leading parton and radiated partons made by the {\em a priori} assumtion that the energy of the leading parton is much larger than the energy of radiation quanta. In this limit, the medium-induced energy loss is given by the sum of the energy of all radiated quanta or equally by the energy difference of the leading parton before and after the passage through the medium. {\em A posteriori}, one may however find (rare) processes in which the energy of a radiated parton is much larger than the energy of the parent --- these are formally counted as large energy loss. 

In a shower model, the leading (most energetic) parton can be identified at all times, but the identity of that parton may change easily during the evolution. Unlike in the energy loss case, the radiation of a hard gluon carrying away 80\% of the energy of a parent quark would not be counted as large energy loss but as change in the identity of the leading parton. No particular branching process can be attributed uniquely to vacuum or medium --- differences between a medium-modified and a vacuum shower can only be defined at the level of averages over many events. This implies that the baseline to compare with is not a single parton with a definite energy, but rather a vacuum shower with a statistical distribution of hadrons at given momenta.

As a way to overcome these difficulties and to make contact between in-medium shower and energy loss modelling, in \cite{YaJEM2} $c$-quarks as shower initiators have been considered. Due to the deadcone effect, these have comparatively low vacuum shower activity and a hard fragmentation pattern, thus minimizing the distortion of the vacuum baseline from a single parton with definite energy. Moreover, since $c$ quark generation in branchings $g\rightarrow c \overline{c}$ is mass suppressed, tagging the $c$ quark almost always tags the shower initiator.

Energy loss of the leading $c$ quark can then be deduced from the ansatz

\begin{equation}
\label{E-Folding}
\frac{dN}{dE}_c^{med}\negthickspace \negthickspace(E) = \int d (\Delta E) \frac{dN}{dE}_c^{vac}\negthickspace\negthickspace(E') P(\Delta E) \delta(E' - E - \Delta E)
\end{equation}

solving for the probability distribution of energy loss $P(\Delta E)$. Note that this ansatz contains the rather drastic assumption that there is no parametric dependence on the initial energy $E$. If we require $P(\Delta E)$ to be a probability distribution, the assumption may imply that for some partons in the distribution $dN/dE_c^{vac}$ the energy loss $\Delta E$ is larger than their energy $E$ in which case they have to be considered lost to the medium. 

In practice, Eq.~(\ref{E-Folding}) is discretized and cast it into the form of a matrix equation

\begin{equation}
\label{E-Matrix}
N_i(E^i) = \sum_{j=1}^n K_{ij}(E^i, \Delta E^j) P_j(\Delta E^j)
\end{equation}

where $dN/dE_c$ is provided at $m$ discrete values of $E$ labelled $N_i$ and $P(\Delta E)$ is probed at $n$ discrete values of $\Delta E$ labelled $P_j$. The kernel $K_{ij}$ is then the calculated $dN/dE_c$ for all pairs $(E^i, \Delta E^j)$ where the energy loss acts as a shift of the distribution, i.e. $dN/dE_c^{med}(E) = dN/dE_c^{vac}(E+\Delta E)$.
Eq.~(\ref{E-Matrix}) can in principle be solved for the vector $P_j$ by inversion of $K_{ij}$ for $m=n$. However, in general this does not guarantee that the result is a probability distribution. Especially in the face of statistical errors and finite numerical accuracy the direct matrix inversion may permit negative $P_j$ which have no probabilistic interpretation.
Thus, a more promising solution which avoids the above problems is to let $m > n$ and find the vector $P$ which minimizes $|| N - K P||^2$ subject to the constraints $0 \le P_i \le 1$ and $\sum_{i=1}^n P_i = 1$. This guarantees that the outcome can be interpreted as a probability distribution and since the system of equations is overdetermined for $m>n$ errors on individual points $R_i$ do not have a critical influence on the outcome any more.

Note that this ansatz leads to additional complications in combination with the prescription to set the minimum virtuality scale $Q_0$ in the shower Eq.~(\ref{E-Q0}). This is due to two reasons:  First, in essence for short paths, energy loss cannot be extracted reliably because the scale choice for the vacuum part becomes only irrelevant if the vacuum shower is a small correction to the medium-induced shower as discussed above. Unfortunately, for truly long paths the second complication becomes apparent: unlike in the light quark case, the charm mass $m_c$ rather than Eq.~(\ref{E-Q0}) determines the point at which the branching is stopped. In order to avoid the second limit, we use very energetic $c$ quarks to extract energy loss, thus allowing for a larger length scale to be probed. 

For these reasons, the actual effect of Eq.~(\ref{E-Q0}) on the pathlength dependence of light parton energy loss is only in a very limited way represented by evaluating Eq.~(\ref{E-Matrix}), and all energy loss results shown for this prescription need to be treated with the appropriate amount of caution. However, this is largely a complication of the way energy loss is determined from an in-medium shower result --- there is no compelling reason to suspect that  Eq.~(\ref{E-Q0}) would be as problematic for the in-medium shower evolution result as such or in the computation of observables. 

As a side remark, the fact that the mass of heavy quarks largely set the scale at which the shower evolution is terminated implies that there is no substantial heavy-quark radiative energy loss in the kinematical region of RHIC experiments in YaJEM. Without an additional mechanism like elastic energy loss, YaJEM thus fails (like other radiative energy loss models) to account for the suppression of heavy quarks seen in single electron spectra. 

\section{Results}

\subsection{Pathlength dependence of energy loss}

\begin{figure}[htb]
\begin{center}
\epsfig{file=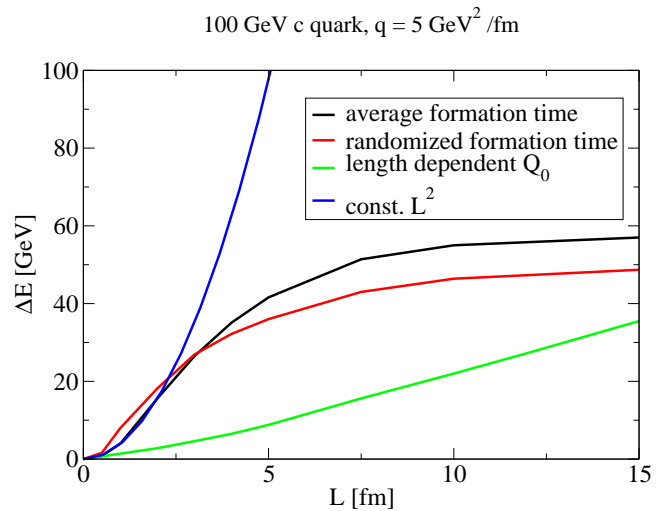, width=8.5cm}
\end{center}
\caption{\label{F-DeltaE}(Color online) The integrated mean energy loss $\Delta E$ as a function of pathlength extracted from YaJEM for a 100 GeV shower-initiating $c$ quark propagating through a constant medium with $\hat{q} = 5$ GeV$^2$/fm for different options (see text), compared with a quadratical dependence on pathlength.}
\end{figure}

In Fig.~\ref{F-DeltaE} the mean energy loss as a function of pathlength extracted from YaJEM via Eq.~(\ref{E-Matrix}) for a 100 GeV charm quark as shower initiator is shown for different options.

In the first case, the formation time of virtual states Eq.~(\ref{E-Lifetime}) is not randomized by Eq.~(\ref{E-RLifetime}). The result is initially very well described by a quadratic length dependence until (around 3 fm) finite energy corrections become important and the limit of applicability is reached and turn the curve over.

If randomization is taken into account the quadratic dependence becomes much blurred, and the net result can be described well by a linear growth with pathlength until finite energy corrections and the limit of applicability turn the result over to a constant again. This agrees nicely with the scaling law identified in \cite{YaJEM2} and shows that an algorithm which in principle reproduces the analytical $L^2$ dependence in the right limit may not show this behaviour in practically relevant situations.

Finally, the last case shows the effect of a variable minimum virtuality scale $Q_0$ according to Eq.~(\ref{E-Q0}). Note that, following the previous discussion, certainly beyond 10 fm the result is not characteristic for what would happen in the light quark case as the charm quark mass rather than $Q_0$ determines where the branching process is terminated. For a light quark, the curve would rise more strongly. 

\subsection{Pathlength dependence of the nuclear suppression factor $R_{AA}$}

In order to illustrate how pathlength effects manifest in observable quantities, we use a realistic 3+1d hydrodynamical evolution model \cite{hydro3d} for the bulk medium and compute the dependence of the nuclear suppression factor $R_{AA}$ on the angle with the reaction plane in this background. The straightforward computation is described in detail in \cite{YaJEM1,Asymptotics}. It involves averaging the medium-modified fragmentation function  (MMFF) computed from YaJEM over all possible paths through the medium, which are in turn determined by the distribution of binary collisions in the transverse plane and the orientation of the parton momentum vector with respect to the reaction plane. This result is then folded with a leading order perturbative Quantum Chromodynamics (LO pQCD) calculation for the parton spectrum to get the yield of high $P_T$ hadrons. Since YaJEM obtains the MMFF at given partonic scale whereas LO pQCD computations require the fragmentation function to be known at given hadronic scale, a scale matching procedure as described in \cite{Asymptotics} has been used in the following.

In all computations, the factor $K$ in Eq.~(\ref{E-qhat}) is adjusted such that the measured $R_{AA}$ for 0-10\% central Au-Au collisions at $P_T = 6$ GeV is reproduced by the calculation. Both the $P_T$ dependence of $R_{AA}$, the centrality dependence and the dependence on the angle $\phi$ with the reaction plane follow then without additional free parameters. Note that the value of $K$ depends on the choice of $Q_0$ (as pointed out already in \cite{YaJEM2}).

\begin{figure}[htb]
\begin{center}
\epsfig{file=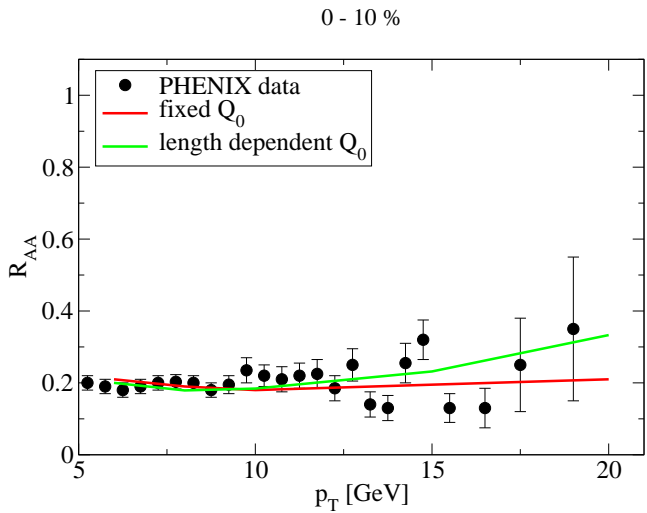, width=8.5cm}
\end{center}
\caption{\label{F-RAA-0-10}(Color online) The nuclear suppression factor $R_{AA}$ for 0-10\% central 200 AGeV Au-Au collisions computed for two different options (see text) in YaJEM, compared with PHENIX data \cite{PHENIX_R_AA}.}
\end{figure}

In Fig.~\ref{F-RAA-0-10} the result for the angular average of $R_{AA}$ is shown in comparison with PHENIX data \cite{PHENIX_R_AA} for both a fixed value of the minimum virtuality scale $Q_0 = 0.7$ GeV and the dynamical choice $Q_0 = \sqrt{E/L}$ where $L$ is the length of the parton path in the medium. As discussed in more detail in \cite{SysHyd3}, there is no compelling reason to identify the Cooper-Frye freeze-out hypersurface in a hydrodynamical model with a sharp surface beyond which the medium interaction with partons is absent, and prescription to let the medium fade out gradually would be desirable. However, since the length dependence enters under the square root, the dependence on details is sufficiently weak such that one may hope to get a reasonable estimate of the effect.

As seen from the figure, both computations are able to describe $R_{AA}$ well, however the $P_T$ dependence is somewhat different. This is not surprising, since the expression for $Q_0$ also contains a dependence on the initial parton energy $E$ which is proportional to the final hadron $P_T$. Generically, for the dynamical scale choice, a rise of $R_{AA}$ with increasing $P_T$ is expected. The curve for fixed $Q_0$ has been obtained for a randomized formation time. The result computed for average formation time is not shown, as it is indistinguishable given the averaging over many different in-medium pathlengths as required in the computation of $R_{AA}$.

Note that the $P_T$ dependence of $R_{AA}$ is different from results shown in \cite{YaJEM1,YaJEM2}. The reason is that in these works the matching procedure between hadronic and partonic scale described in \cite{Asymptotics} has not been used, which leads to an unrealistic {\em decrease} of $R_{AA}$ with $P_T$.

\begin{figure}[htb]
\begin{center}
\epsfig{file=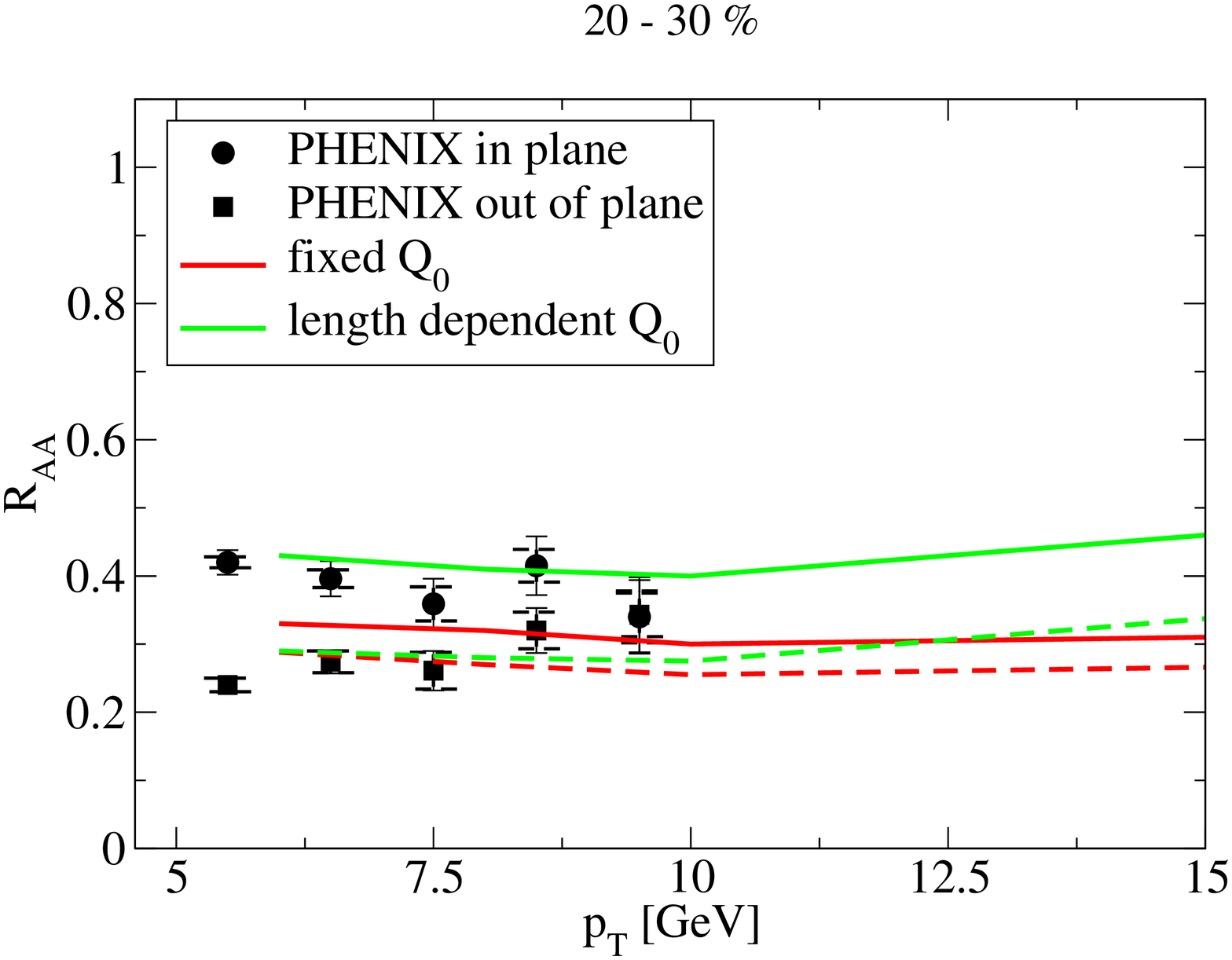, width=8.5cm}
\end{center}
\caption{\label{F-RAA-20-30}(Color online) The nuclear suppression factor $R_{AA}$ for 20-30\% central 200 AGeV Au-Au collisions computed for two different options (see text) in YaJEM, compared with PHENIX data \cite{PHENIX_RAA_phi}. Shown are in-plane emission (solid) and out of plane emission (dashed). }
\end{figure}

The result for 20-30\% centrality is shown in Fig.~\ref{F-RAA-20-30}, separated into in plane and out of plane emission, again for a fixed value of $Q_0$ and a dynamically computed value and compared with PHENIX data \cite{PHENIX_RAA_phi}.

It is immediately apparent from the figure that a computation with fixed $Q_0$ results in a spread between in-plane and out of plane emission much smaller than seen in the data. This is consistent with the observation made earlier that the effective pathlength dependence of energy loss in YaJEM is approximately linear, and the spread found agrees with results from models in which the dependence was assumed to be linear \cite{Elastic1,Elastic2}. 

On the other hand, both spread and normalization of the data are well reproduced by a dynamical choice for $Q_0$ which introduces for light quarks and gluons a strong pathlength dependence of energy loss. 

The scenario with dynamically computed $Q_0$ has thus clear observable consequences when observing hadrons at higher energies, such as available at the LHC. Since the maximal in-medium pathlength is always limited by the nuclear diameter times a factor of order one, Eq.~(\ref{E-Q0}) implies that $Q_0$ will grow with increasing energy, thus the suppression induced by the medium will decrease, leading to a rise of $R_{AA}$ with increasing $P_T$. Somewhat unfortunately, at LHC kinematics, such a rise is expected in various models for a number of reasons, among them the shape of the parton spectrum or the fact that the MMFF is, even for fixed $Q_0$, energy dependent and reverts for large energies to the vacuum result \cite{Asymptotics}. However, the rise induced by the dynamically computed $Q_0$ occurs much faster than any of these other effects.

In contrast, the ratio of computed $Q_0$ at fixed energy for different pathlengths is a constant. When considering in-plane and out of plane emission for non-central collisions, this implies that while the mean (angular averaged) $R_{AA}$ changes rapidly with energy, the spread between in-plane and out of plane will be more robust and change much slower with energy (since asymptotically at high energies the MMFF reverts to the vacuum FF, eventually the spread must vanish as well). Thus, the scenario proposed here can be tested by an observation of in-plane vs. out of plane emission of hard hadrons in non-central collisions for a sufficiently large range in $P_T$. Detailed predictions for LHC kinematics will be postponed to a subsequent publication, as they involve a significant numerical effort.

\section{Discussion}

From the results of the previous section, it would appear that in the kinematical domain currently experimentally accessible, imposing a global virtuality resolution scale Eq.~(\ref{E-Q0}) down to which the medium can influence the shower development has a much larger influence on the observable pathlength dependence of high $P_T$ hadron production than a local coherence condition imposed for each branching which accounts for LPM interference. 

Note that $Q_0$ acts in the presence of a medium quite different than in the vacuum case. In vacuum shower, $Q_0$ regulates the transition from a perturbative partonic description in terms of a shower evolution to a non-perturbative hadronization description in terms of the Lund model, and there is some transition region in which both prescriptions lead to similar results such that the final result is not strongly influenced by the detailed scale choice. However within YaJEM, by construction (and due to a lack of insight into what physics hadronization inside the medium would entail), the medium only influences the partonic dynamics and there is no compensation of e.g. a high scale choice leading to weak showering by an increased action of the Lund model as in the vacuum case. Thus, here $Q_0$ more directly regulates what part of the shower evolution happens in medium and what needs to be taken care off by vacuum dynamics later.  

If this result is confirmed by more detailed investigations, it would imply the the idea to use $R_{AA}(\phi)$ to distinguish between pQCD and strongly coupled models on the basis of quadratic vs. cubic pathlength dependence is largely a red herring, as the leading effect seen in the data is something entirely different. However, since the scenario also predicts an energy dependence of the virtuality scale, translating e.g. into a rise of $R_{AA}$ with $P_T$, testing it should be comparatively easy with a large enough kinematic lever-arm as e.g. the LHC experiments can provide. 

In general, it seems that in the context of MC shower models which can incorporate finite energy effects and to some degree also quantum interference, the question of how pathlength dependence arises and how to define it from observables becomes significantly more complicated than in the asymptotic limits of analytical models.

\begin{acknowledgments}
Discussions with Abhijit Majumder, Korinna Zapp and Guangyou Qin are gratefully acknowledged. This work was 
supported by an Academy Research Fellowship of T.R. from the Finnish Academy (Project 130472) and 
from Academy Project 133005.
 
\end{acknowledgments}

\end{document}